\documentclass[11pt]{article}
\usepackage[dvipdfmx]{graphicx}
\usepackage{color}
\begin{document}

\begin{center}{\Large Information Criterion for a Large Scale Subset Regression Models}\\

\vspace{8mm}
{\large Genshiro Kitagawa}\\[2mm]
The Institute of Statistical Mathematics\\[-1mm]
and\\[-1mm]
Graduate University for Advanced Study

\vspace{3mm}
{\today}

\end{center}

\noindent{\bf Abstract}

The information criterion for determining the number of explanatory variables in a subset regression modeling is discussed. Information criterion such as AIC is effective and frequently used in model selection for ordinary regression models and statistical models. With the recent prosperity of data science, analysis of large-scale data has become important. When constructing models heuristically from a very large number of candidate explanatory variables, there is a possibility of picking up apparent correlations and adopting inappropriate variables. In this paper, we point out the problems specific to subset regression from the viewpoint of bias correction for log-likelihood and present a correction method that takes this into account.

\vspace{1mm}
\noindent

\section{Introduction}


Developments in data science have increased the need to analyze large data sets.
For example, the estimation of regression models has traditionally used the information criterion AIC or a modified version of it to select variables that are effective in explaining or predicting variation in the objective variable from a relatively small number of carefully designed and obtained explanatory variables.

In recent years, however, the number of candidate explanatory variables is inexhaustible, and finding useful variables among them to predict explanatory variables has become an important issue. This is because it is possible to find useful variables other than the candidate explanatory variables selected based on the analyst's knowledge. However, the computational difficulties involved in estimating subset regression models when there are many explanatory variables make the problem of variable selection even more serious. Conventional least-squares estimation and order selection result in the selection of a larger number of variables than expected, and in recent years, estimation using the $L_1$ criterion and other criteria has become the popular. In this case, however, the cross-validation method is often used for parameter selection, which is likely to require a great deal of time for evaluation in the case of very large data
In this paper, we consider the information criterion when estimating subset regression models in situations where only a relatively small number of variables are actually valid among a large number of explanatory variables.

Many methods have been developed for variable selection in multiple regression analysis.
Mantel (l970) and Beale (l970) examined the efficiency of this procedure in terms of computational effort and desirability of the final results.
However, they relied exclusively on the sum of squared residuals as a criterion for comparing fitted models.

In statistical inference situations, Akaike (l973) proposed the consistent use of entropy:
\begin{eqnarray}
B(f;g) = \int \log \left\{\frac{g(z;x)}{f(z)}\right\} f(z)dz,
\end{eqnarray}
where $x$ is the vector of observations and $f(z)$ and $g(z;x)$ are the probability density functions of the true and estimated models. According to the entropy maximization principle, the objective of statistical inference is to estimate $f(x)$ from the data $x$ and find $g(z;x)$ that maximizes the expected entropy $\int B(f;g)f(x)dx$  (or equivalently, the expected log-likelihood).
To achieve this, Akaike (1973, 1974) proposed the information criterion AIC.
AIC corrects for the fact that the log-likelihood has a bias proportional to the number of parameters as an estimate of the expected log-likelihood.
With this correction, AIC has been successfully modeled in various fields using AIC.
However, in the case of subset regression models, special care is needed in evaluating the bias, and it can be seen that well-known criteria such as AIC tend to adopt unnecessary large numbeer of variables.

In this paper, we revisit the method of subset regression from this perspective and attempt to derive model evaluation criteria suitable for this problem.
It will be shown that at each stage of variable selection, the model with the smallest residual sum of squares achieves the smallest expected entropy (or largest expected log-likelihood). In this situation, it is important to recognize that the most efficient procedure for subset regression will select the worst model in terms of entropyexpected log-likelihood at each stage of selection.

The bias correction term in the information criterion AIC is obtained from the expected value of the difference between the log-likelihood and the expected log-likelihood, which can be approximated by a $\chi^2$ distribution. However, in the estimation of subset regression models, since the model with the maximum log-likelihood is selected among many model candidates, it is more appropriate to use the expected value of the extreme value distribution of the $\chi^2$ distribution instead of simply the expected value of the $\chi^2$ distribution. In this paper, we propose specific model selection criterion based on our considerations under a variety of conditions.

\section{A Review of the Subset Regression Technique}

Assume that $N$ observations of $(k+1)$-dimensional vector ($y_n,x_{1n},\ldots ,x_{kn}$) are given, where $y_n$ is the objective variable and $x_{1n},\ldots ,x_{kn}$ are $k$-dimensional explanatory variables. 
For simplixity, we assume that the
explanatory variables are orthogonal and satisfy $E [x_{in}] = 0$, 
$E [x_{in}^2] =1$ and $E [x_{in}x_{jn}] =0$ for $i \neq j$.

We consider a linear regression model
\begin{eqnarray}
  y_n = \sum_{i=1}^k a_i x_{in} + \varepsilon_n , \label{Eq_full-order model}
\end{eqnarray}
where $a_i$ is the regression coefficient and $\varepsilon_n$ is 
a Gaussian random variable with mean 0 and variance $\sigma^2$. 
The parameter vector is defined by $\theta = (a_1,\ldots ,a_k,\sigma^2)$
and the true parameter is denoted as $\theta^* = (a_1^{*},\ldots ,a_k^{*},\sigma^2_{*})$. It is also assumed that $a_j^* =0$ for $j \geq k^*$. If $k^{\ast} \ll k $, 
it indicates that only a few variables out of a large number of explanatory variables actually have information useful for predicting the objective variable.

For big data situation where $k$ is very large, it is not practical
to estimate the full-order model (\ref{Eq_full-order model}), and
consider the subset regression model
\begin{eqnarray}
  y_n = \sum_{j=1}^m a_{i_j} x_{i_jn} + \varepsilon_n ,\qquad (m \ll k).
\end{eqnarray}
Here $m$ is referred to as the order of the subset regression model, and $I_m=(i_1,\ldots,i_m)$ indicates the explanatory variable used
in subset regression model.
The parameter vector is defined by $\theta_{m}(I_m) = (a_{i_1},\ldots ,a_{i_m},\sigma_m^2)$.

Given data $\{ (y_n,x_{in},i=1,\ldots,k),n=1,\ldots ,N\}$,
the log-likelihood of the subset regression model is given by
\begin{eqnarray}
\ell (\theta_{m}(I_m))
= -\frac{N}{2}\log 2\pi \sigma_m^2 - \frac{1}{2\sigma_m^2}\sum_{n=1}^N \biggl( y_n - \sum_{j=1}^m a_{i_j}x_{i_jn}\biggr)^2.
\end{eqnarray}
The maximum likelihood estimate of the regression coefficients and the residual variance of the model are respectively obtained by
\begin{eqnarray}
  \hat{a}_{i_j} = \biggl(\sum_{n=1}^N x_{i_jn}^2\biggr)^{-1}{\sum_{n=1}^N y_nx_{i_jn}},
\end{eqnarray}
and
\begin{eqnarray}
  \hat{\sigma}_m^2(i_1,\ldots ,i_m) 
= N^{-1}\biggl(\sum_{n=1}^N y_n^2 - \sum_{j=1}^m \hat{a}_{i_j}^2 \sum_{n=1}^N x_{i_jn}^2 \biggr) = N^{-1}\biggl(\sum_{n=1}^N y_n^2 - \sum_{j=1}^m \hat{a}_{i_j}^2 \biggr).
\end{eqnarray}
The maximum log-likelihood and the AIC are given by
\begin{eqnarray}
\ell (\hat{\theta}_m(\hat{I}_m))
&=& -\frac{N}{2}\log 2\pi \hat{\sigma}_m^2 - \frac{N}{2}, \\
\mbox{AIC}_m &=& -2 \ell (\hat{\theta}_m(\hat{I}_m)) + 2(m+1) \nonumber \\
  &=& N \log 2\pi \hat{\sigma}_m^2 + N + 2(m+1).
\end{eqnarray}

The entropy of the true model $q(y|\theta^*)$ with respect to the fitted subset regression model $p(y|\hat{\theta}_m)$ is obtained by 
\begin{eqnarray}
I( q(y|\theta^*); p(i|\hat{\theta}_m))
&=& \int \log \frac{p(y|\hat{\theta}_m)}{q(y_m|\theta^*)} q(y|\theta^*)dy \nonumber \\
&=& E_q\Bigl[ \log p(y|\hat{\theta}_m ) \Bigr] - E_q\Bigl[ \log q(y|\theta^* ) \Bigr],
\end{eqnarray}
where $E_q\Bigl[ \log p(y|\hat{\theta}_m ) \Bigr]$ and $E_q\Bigl[ \log q(y|\theta^* ) \Bigr]$ are obtained by
\begin{eqnarray}
 E_q\Bigl[ \log p(y|\hat{\theta}_m ) \Bigr]
 &=& -\frac{1}{2} \log 2\pi \hat{\sigma}^2_m -\frac{1}{2\hat{\sigma}^2_m}
      E_q \biggl[ \biggl(y - \sum_{j=1}^k \hat{a}_j x_{jn}\biggr)^2 \biggr] \nonumber \\
 &=& -\frac{1}{2} \log 2\pi \hat{\sigma}^2_m -\frac{1}{2\hat{\sigma}^2_m}
      E_q \biggl[ \biggl\{ y - \sum_{j=1}^k a^*_j x_{jn} + \sum_{j=1}^k (a^*_j - \hat{a}_j) x_{jn} \biggr\}^2 \biggr] \nonumber \\
 &=& -\frac{1}{2} \log 2\pi \hat{\sigma}^2_m - \frac{1}{2\hat{\sigma}^2_m}
      \biggl\{ \sigma_*^2  + \sum_{j=1}^k (a^*_j - \hat{a}_j)^2  \biggr\} \nonumber \\
 E_q\Bigl[ \log q(y|\theta^* ) \Bigr]
 &=& -\frac{1}{2} \log 2\pi \sigma^2_* - \frac{1}{2}.
\end{eqnarray}
Therefore the entropy of the true model with respect to the estimated subset
regression model is obtained by
\begin{eqnarray}
I( q(y|\theta^*); p(y|\hat{\theta}_m))
 &=& E_q\Bigl[ \log p(y|\hat{\theta}_m ) \Bigr] - E_q\Bigl[ \log q(y|\theta^* ) \Bigr] \nonumber \\ &=& \frac{1}{2} \biggl[ \log \frac{\sigma^2_*}{\hat{\sigma}^2_m} + 1
    - \frac{1}{\hat{\sigma}^2_m}
      \biggl\{ \sigma_*^2  + \sum_{j=1}^k (a^*_j - \hat{a}_j)^2  \biggr\} \biggr]
\end{eqnarray}
where  $\displaystyle\sum_{j=1}^k (a^*_j - \hat{a}_j)^2 $ can be decomposed into two terms as
\begin{eqnarray}
  \sum_{j=1}^k (a^*_j - \hat{a}_j)^2  
  &=& \sum_{j\in I_m} (a^*_j - \hat{a}_j)^2  + \sum_{j\not\in I_m} (a^*_j)^2 \nonumber \\
  &=& \sum_{j\in I_m} (e_j)^2  + \sum_{j\not\in I_m} (a^*_j)^2,
\end{eqnarray}
where $e_j = a^*_j - \hat{a}_j$ and $\hat{a}_j$ is given by
\begin{eqnarray}
 \hat{a}_j \sim N(a_j^*,\sigma^2_*(X^TX)^{-1}) = N(a_j^*,N^{-1}\sigma^2_*). \label{Eq_(11)}
\end{eqnarray}

\section{A Special Case: Independent True Model Case}
To specifically evaluate the entropy obtained in the previous section, we first consider the special case where all true coefficients are zero, i.e., we assume that $a^*_j=0$, $j=1,\ldots ,k$; 
\begin{eqnarray}
  y_n = \varepsilon_n, \quad \varepsilon_n \sim N(0,\sigma^2_*).
\end{eqnarray}
Here, we considere the following subset regression model
\begin{eqnarray}
  y_n = \sum_{j=1}^m a_{i_j}x_{i_jn} + \varepsilon_n, \quad \varepsilon_n \sim N(0,\sigma^2).
\end{eqnarray}
Given $N$ observations, the parameter vector is estimated as $\hat{\theta}_m = (\hat{a}_{i_1},\ldots ,\hat{a}_{i_m},\hat{\sigma}^2) $, where the regression coefficient satisfies 
\begin{eqnarray}
 \hat{a}_{i_j} &\sim& N(0,N^{-1}\sigma_*^2). \nonumber 
\end{eqnarray}
Here, since there are $k$ candidates of regressors and the true coefficients are assumed to be zero, the term in equation (\ref{Eq_(11)}) is expressed as the $\chi^2$ random variable with $m$ degrees of freedom;
\begin{eqnarray} 
 \sum_{j=1}^k (a^*_j - \hat{a}_j)^2 &=& \sum_{j\in I_m}\hat{a}_j^2  \sim  \chi^2_m .
\end{eqnarray}
However, the estimation algorithm for the subset regression model selects the combination that is the largest value among such a large number of $\chi^2_m$ variables.
To obtain the expectation of extreme value of $\chi^2_m$ distribution in $L=\left( \begin{array}{c} k \\ m \end{array}\right)$ possible candidate subset regression models, 
define the ordered statistics of the $\chi^2_m$ random variable in order of magnitude
\begin{eqnarray}
  X_{(1)} \geq X_{(2)} \geq \ldots \geq X_{(L)}.
\end{eqnarray}
If it is necessary to explicitly state that it is the $r$-th order among $N$ variables, we will denote it as $X_{(r|N)}$. Under the current assumption that all explanatory variables are orthogonal, the expected value of this extreme value can be expressed by summing the ordinal statistics of the chi-square variables with one degree of freedom as 
\begin{eqnarray}
 \sum_{j=1}^r E[X_{(j|N)}],
\end{eqnarray} 
where the distribution of the $r$-th ordered statistics and the extreme value are respectively given by
\begin{eqnarray}
  p_{x_{(r)}}(x) &=& \frac{N!}{(N-r)!(r-1)!}F(x;1)^{N-r}\{1-F(x;1)\}^{r-1}f(x;1) \\
  p_{x_{(1)}}(x) &=& N F(x;1)^{N-1}f(x;1),
\end{eqnarray}
where the density function $f(x;m)$ and the cummulative distribution function $F(x;1)$ are respectively given by
\begin{eqnarray}
f(x;1) &=& \frac{1}{2^{1/2}\Gamma (\frac{1}{2})}x^{-\frac{1}{2}}\exp\left\{-\frac{x}{2}\right\} \\
F(x;1) &=& \frac{\gamma(\frac{1}{2},\frac{x}{2})}{\Gamma (\frac{1}{2})}. \label{Eq_incomplete_Gamma}
\end{eqnarray}
Here $\Gamma (\frac{1}{2})= \sqrt{\pi}$ and  $\gamma(\frac{1}{2},\frac{n}{2})$ is the incomplete Gamma function which given by
\begin{eqnarray}
\gamma(1/2,x) &=& \sqrt{\pi} \mbox{erf}(\sqrt{x}), \nonumber 
\end{eqnarray}
where $\mbox{erf}(x)$ is the error function defined by
\begin{eqnarray}
\mbox{erf}(x) = \frac{2}{\sqrt{\pi}}\int_0^x \exp^{-t^2} dt.
\end{eqnarray}

Since the regression variables are mutually independent, the expected value of the sum of $k$ extreme values can be obtained as the sum of the expected values,
$E[X_{(1|N)}]+\cdots +E[X_{(r|N)}]$.
The expected value of the $X_{(1)}$ and $X_{(r)}$ are obtained by
\begin{eqnarray}
  E[ X_{(1|N)}] &=& N \int_{x=0}^{\infty} x F(x;k)^{N-1} f(x;k) dx, \nonumber \\
  E[ X_{(r|N)}] &=& \frac{N!}{(N-r)!(r-1)!} \int_{x=0}^{\infty} x F(x;k)^{N-r}(1-F(x;k)^{r-1} f(x;k) dx . 
\end{eqnarray}

\begin{figure}[tbp]
\begin{center}
\includegraphics[width=75mm,angle=0,clip=]{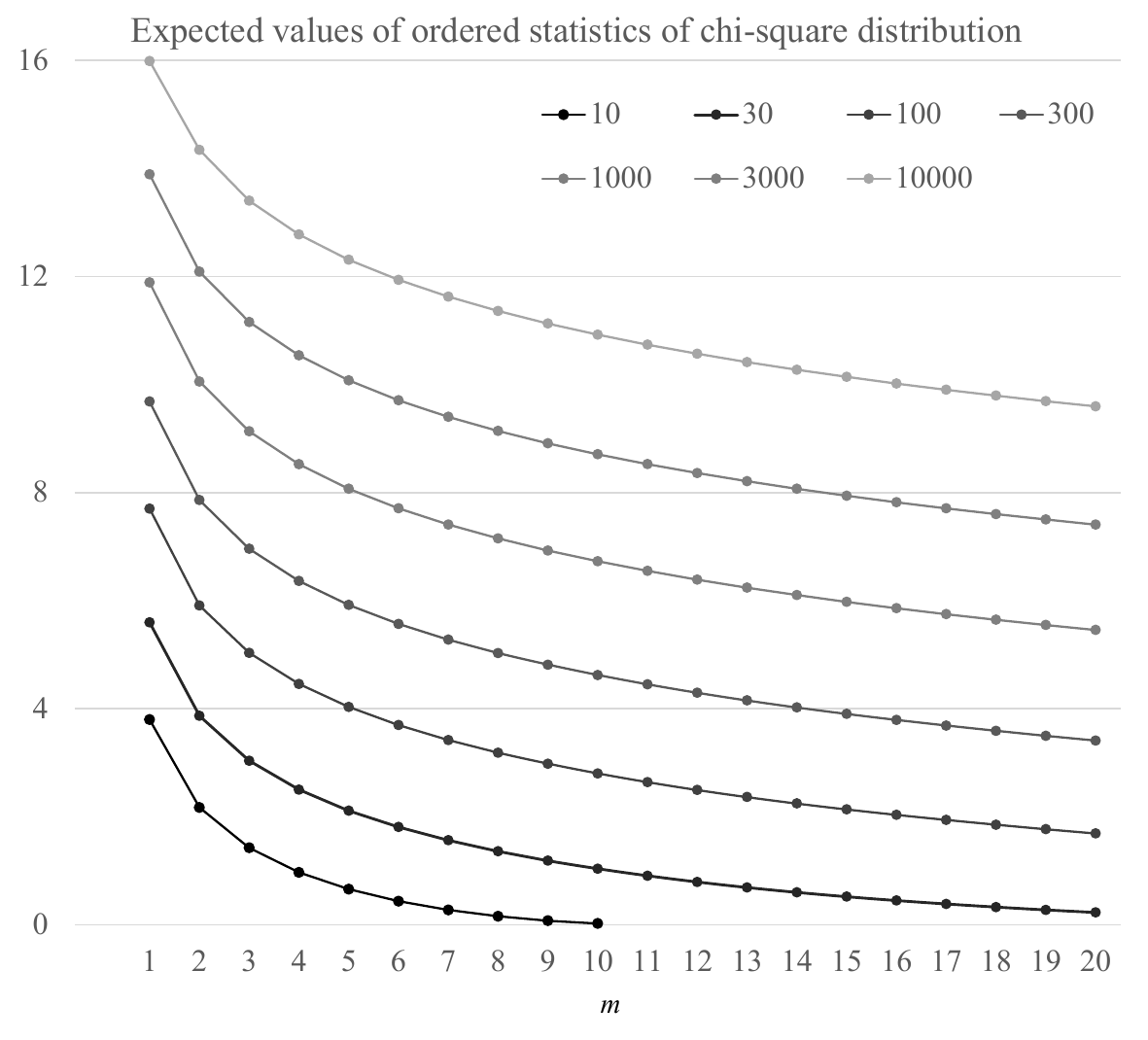}
\hspace{4mm}
\includegraphics[width=75mm,angle=0,clip=]{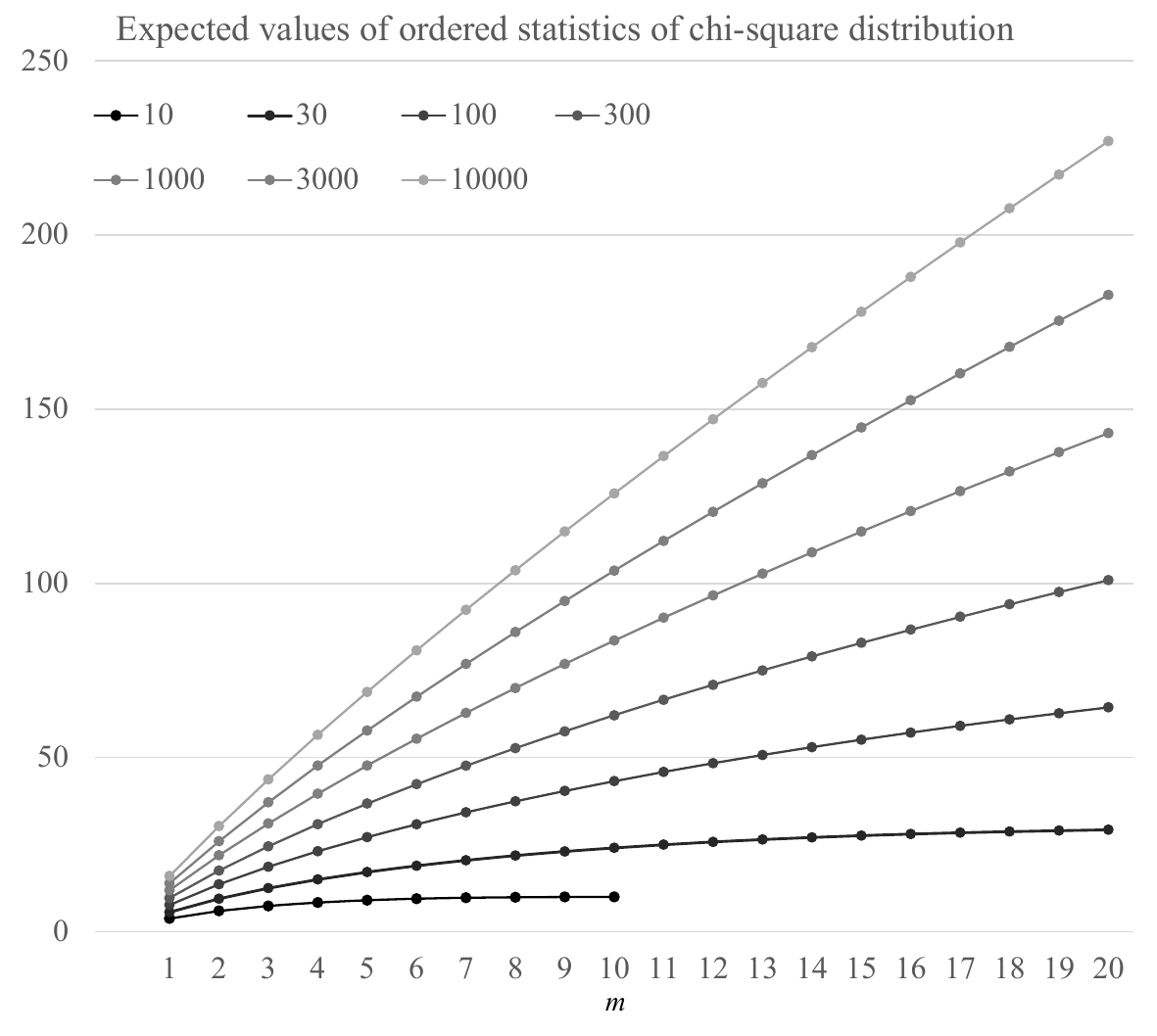}
\end{center}
\caption{Expected values of the sum of $k$ extreme values of $\chi^2$ random variables of degree one. From top to bottom, $k$=10,30,100,300,1000,3000,10000. Horizontal axis: order $r$.}
\label{Fig_ordered-chi-square_k=10,30,100,300,1000,3000,10000}
\end{figure}

Left plot of Figure \ref{Fig_ordered-chi-square_k=10,30,100,300,1000,3000,10000} shows the expected values from the first to 20-th largest extreme value statistics when there are $k$ random variables ($k$=10, 30, 100, 300, 1000, 3000, 10000) that follow a $\chi^2$ distribution with one degree of freedom. The corresponding numerical values in this figure are given in Table \ref{Tab_Expacted values of ordered chi-square} in the Appendix. Note that the expected value of the $\chi^2$ distribution with 1 degree of freedom, that corresponds to the AIC, is 1. 

It can be seen from the figure, that even when there are only 10 possible candidates of regressors in subset regression model, the expected value of the extreme values, $E[X_{(1)}]$, becomes 3.80 and it becomes 7.70 for $k=100$, 11.89 for $k=1,000$ and 15.99 for $k=10,000$. The sum of the expected values of all ordered statistics is equal to $k$ times the expected value of $\chi^2_1$ random variable, i.e., $E[X_{(1)}]+\cdots +E[X_{(k)}]=k$. This means that as the order $r$ increases, the expected value of $r$-th ordered statistics $E[X_{(r)}]$  becomes less than 1 after a certain order $r$. In fact, for $k=10$, $E[X_{(r)}]$ is less than 1 for $r$ greater than 4. Also, even for $k=30$, it is less than 1 for $r$ greater than 11. However, for larger values of $k$, $E[X_{(20)}]$ is much larger than 1 (1,69, 3.41 and 5.46 for $k$=100, 300 and 1000, respectively). Numerical values of this figure is shown in Table \ref{Tab_Expacted values of ordered chi-square} in the appendix.

The right plot of Figure \ref{Fig_ordered-chi-square_k=10,30,100,300,1000,3000,10000} shows their cumurative values. For large $k$, we see that the difference between the expected log-likelihood and the maximum log-likelihood increases almost linearly for $r$ up to 20, and the difference between the expected log-likelihood and the maximum log-likelihood is very large.

\begin{figure}[tbp]
\begin{center}
\includegraphics[width=75mm,angle=0,clip=]{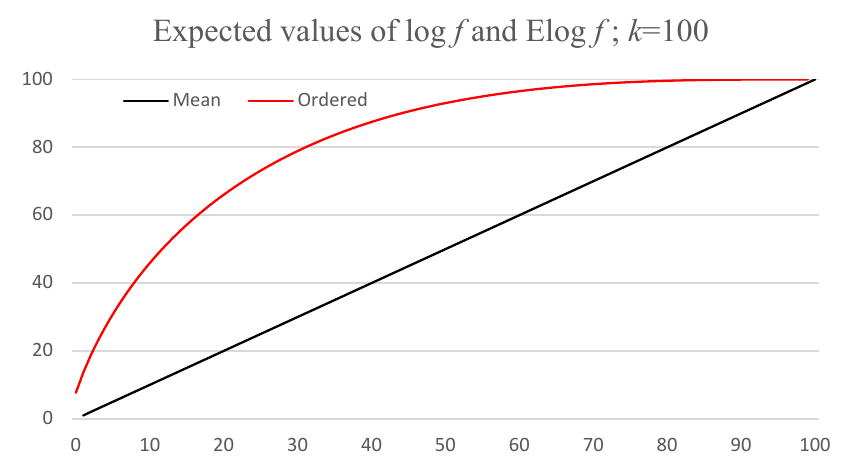}
\hspace{4mm}
\includegraphics[width=75mm,angle=0,clip=]{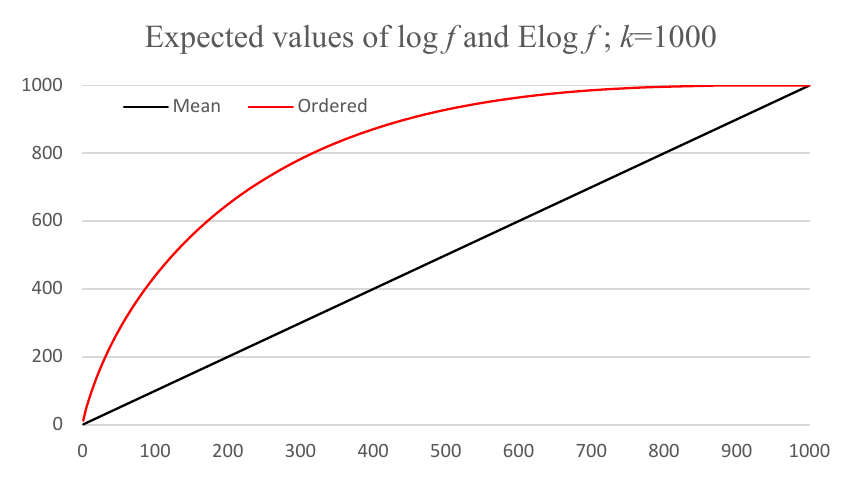}
\end{center}
\caption{Expected values of the sum of $k$ extreme values of $\chi^2$ random variables of degree one. Left plot: $k$=100, Right plot: $k=$1000. Horizontal axis: $r=1,\ldots ,k$.}
\label{Fig_ordered-chi-square_k=100,1000}
\end{figure}

For reference, Figure \ref{Fig_ordered-chi-square_k=100,1000} shows expected values up to the $k$-th ordered statistics are shown for $k$=100 and 1000, respectively. 
In each figure, the red curve indicates the change in the cumulative expected value of the ordered statistics. As the order $r$ approaches the maximum order $k$, the increasing rate of the curve decreases step by step and converges to 0.
The black curve shows the cumulative expected value of the $\chi^2_1$ variable. Since the expected value of the $\chi^2_1$ variable is 1, it is a straight line with constant slope. Naturally, at the right end (i.e., $r=k$), the red and black curves have the same value.



These results indicate that the penalty term of $2\times (20+1)$ for the model with 20 regression coefficients with the AIC criterion is insufficient for selecting relatively low order subset regression model.
%
%
The conservative assumption that all regression coefficients are zero suggests using the following criterion
\begin{eqnarray}
\mbox{SRIC}(m) = -2 \log f(x|\hat{\theta}) + 2\left( \sum_{r=1}^{m} E(X_{(r)}) + 1\right),
\end{eqnarray} 
instead of AIC as the information criterion for order selection in the subset regression model with $r$ regression coefficients. In this criterion, $+1$ represents the penalty corresponding to the estimated error variance of the model.

However, it is more realistic to assume at least some variables are considered significant. In the next section, we consider evaluation of bias under some specific assumption on the true parameters of the model.

\section{Evaluation of Bias in Some Subset Regression Modeling}

In this section, we conider the evalutaion of the bias of the maximum log-likelihood as an estimate of expected log-likelihood, when the true parameters of the regression model are not necessarily zero. The amount of the bias can be evaluated if the effective variables and the true values of their coefficients are known, but it is not appropriate to assume that such information is available in the estimation of the subset regression model.
Therefore, in this section, we assume several patterns of true regression coefficients and calculate the difference between the log-likelihood and the expected log-likelihood.

In this section, we assume that the true model generating the data is given by
\begin{eqnarray}
y_n = \sum_{i=1}^k a_i^{\ast} x_{in} + \varepsilon_n,\quad \varepsilon_n \sim N(0,0.1). \label{Eq_simulation_model}
\end{eqnarray}

\begin{figure}[tbp]
\begin{center}
\includegraphics[width=150mm,angle=0,clip=]{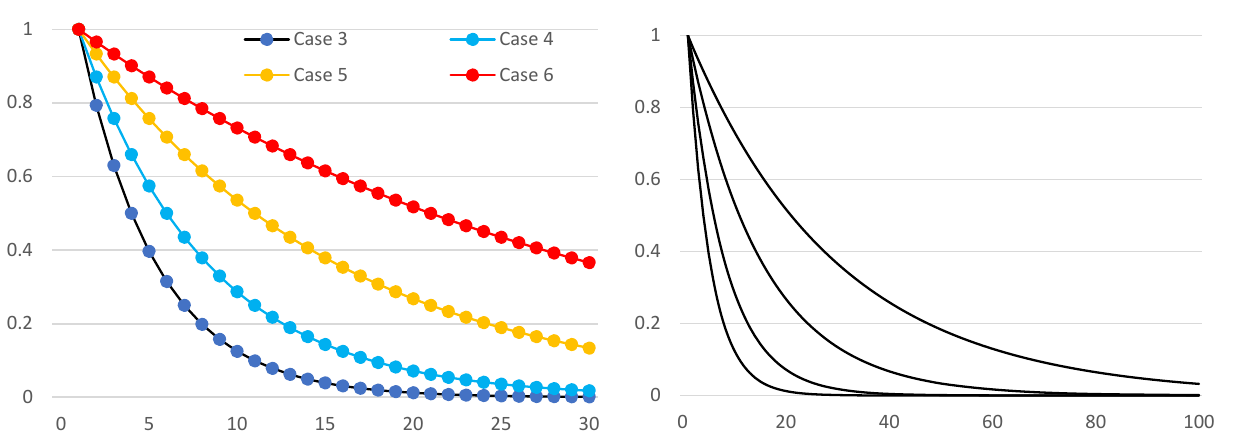}
\end{center}
\caption{Assumed regression models. Left: $j=1,\ldots ,20$, Right: $j=1,\ldots ,100$}
\label{Fig_Assumed model}
\end{figure}

We consider the following six cases:
\begin{enumerate}
\item $a_i^{\ast}=0$ for all $i$ (considered in the previous section).
\item $a_i^{\ast}= 0.8^{i-1}$ for $i=1,\ldots ,10$ and =0 for $i=21,\ldots ,k$.
\item $a_i^{\ast}= C_1^{i-1}$
\item $a_i^{\ast}= C_2^{i-1}$
\item $a_i^{\ast}= C_3^{i-1}$
\item $a_i^{\ast}= C_4^{i-1}$
\end{enumerate}
where $C_j = \exp\{-(\log 2)/m_j\}$ and $m_1=3$, $m_2=5$, $m_3=10$ and $m_4=20$. 
That is, the magnitude of the regression coefficient is halved when $i$ increases by 3, 5, 10, and 20, respectively. Figure \ref{Fig_Assumed model} shows the pattern of true regression coefficients assumed by Case 3 $\sim$ Case 6. Case 2 is almost identical to Case 3, but is a low order model with $a^{\ast}_j$=0 for $j$=11 and above.

For these six cases, $N$ observations  $y_n$ are generated according to the model (\ref{Eq_simulation_model}), where $\varepsilon_n$ is a pseudo-random number with mean 0 and variance 0.1. The explanatory variable $x_{in}$ is a trigonometric function, and an orthogonal functional form is generated. The number of data $N$ was set to 1,000 for $k=30$, 100 and 300, and 10,000 for $k=1000$

\begin{figure}[tbp]
\begin{center}
\includegraphics[width=160mm,angle=0,clip=]{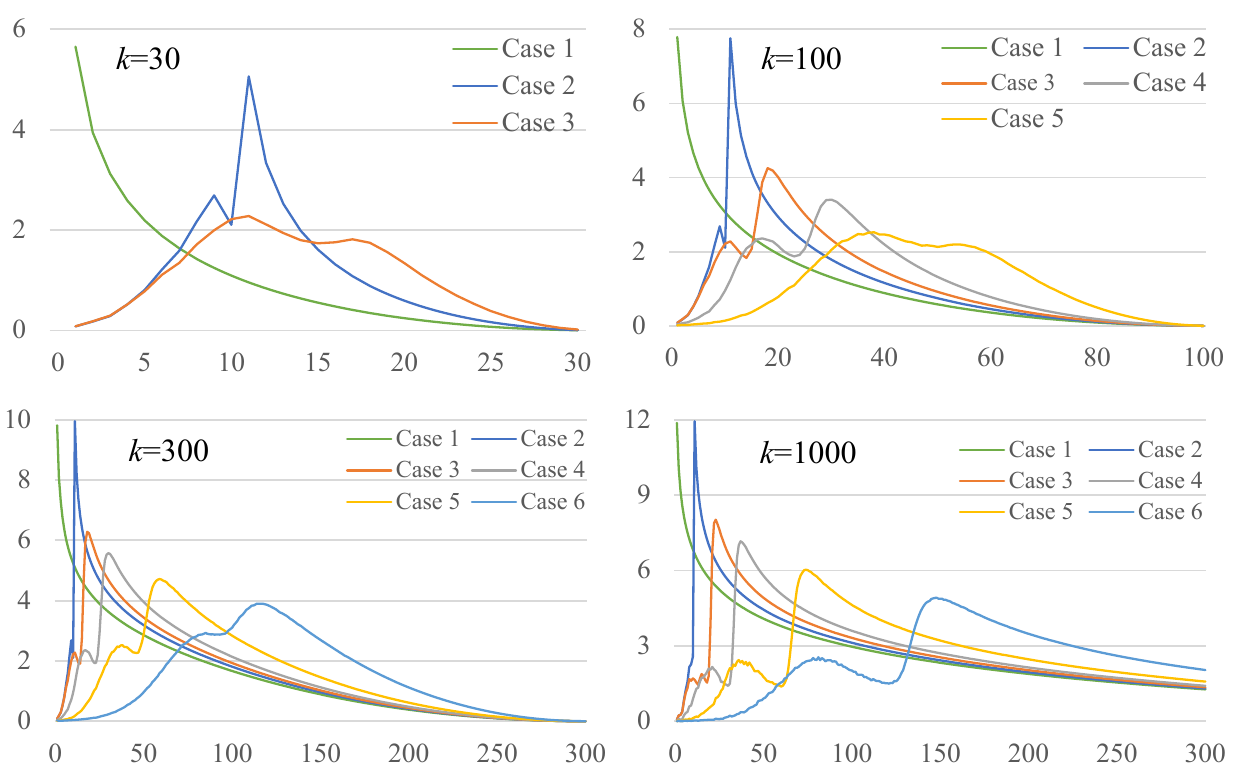}
\end{center}
\caption{Average difference between maximum log-likehood and the expected log-likelihood. Left column: $m=1,\ldots ,30$, Right column: $m=1,\ldots ,k$. From top to bot, $k=30$, 100, 300 and 1000.}
\label{Fig_summary_Case1-6}
\end{figure}

Figure \ref{Fig_summary_Case1-6} shows the difference between the log-likelihood and the expected log-likelihood of the subset regression models for six cases. Subset regression models of each order was estimated for each generated data and the difference between its maximum log-likelihood and expected log-likelihood were calculated. This operation was performed NS times, and the mean value is shown in the figure, where NS is 100,000 times for $k=30$, 100 and 1,000, and 10,000 times for $k=1000$.

Upper left plot of Figure \ref{Fig_summary_Case1-6} shows the  case when the number of candidate regressors is 30. Only the results for Case 1 to Case 3 are shown. In Case 1, where all coefficients are 0, shown by the green curve, the bias starts with a maximum value 5.62 at $j=1$, and becomes less than 1.0 at $j=11$, and then decreases monotonically. In contrast, in Case 2 shown by blue curve, where only the first 10 true coefficients are nonzero, the bias reaches a maximum of about 5.06 at $j=11$ and then declines rapidly. In Case 3, it is not as steep as in Case 2, but reaches a maximum of about 2.27 at $j=11$ and then declines gradually.

The plot in the upper right shows the result for $k=100$. For Case 1, it takes a value 7.79 for $j=1$ and then decays exponentially to 0. Case 2 takes about the same value of 7.76 at $j=11$, and then decays monotonically, as in case 1. In Case 3 and Case 4, the coefficients have one maximum and then monotonically decrease after taking maximum values of 4.25 and 3.40 at $j=18$ and 30, respectively. In Case 5, the maximum value of 2.52 was reached at $j=38$, followed by a gradual decrease.

The lower left plot is the result for $k=300$. In Case 1, the maximum value is 9.82 for $j=1$, and in Case 2, the maximum value is almost the same for $j=11$. In Cases 3-6, the maximum order $j$ shifts to the right and its maximum value becomes smaller and smaller. The bias is monotonically decreasing for all values of $j$ larger than the maximum value.

The lower right plot shows the result for $k=1000$. The result in this case is similar to the case $k=300$, but the order that attains the maximum value of the bias is larger, as well as the maximum value is larger than in the case $k=300$.

\begin{figure}[tbp]
\begin{center}
\includegraphics[width=160mm,angle=0,clip=]{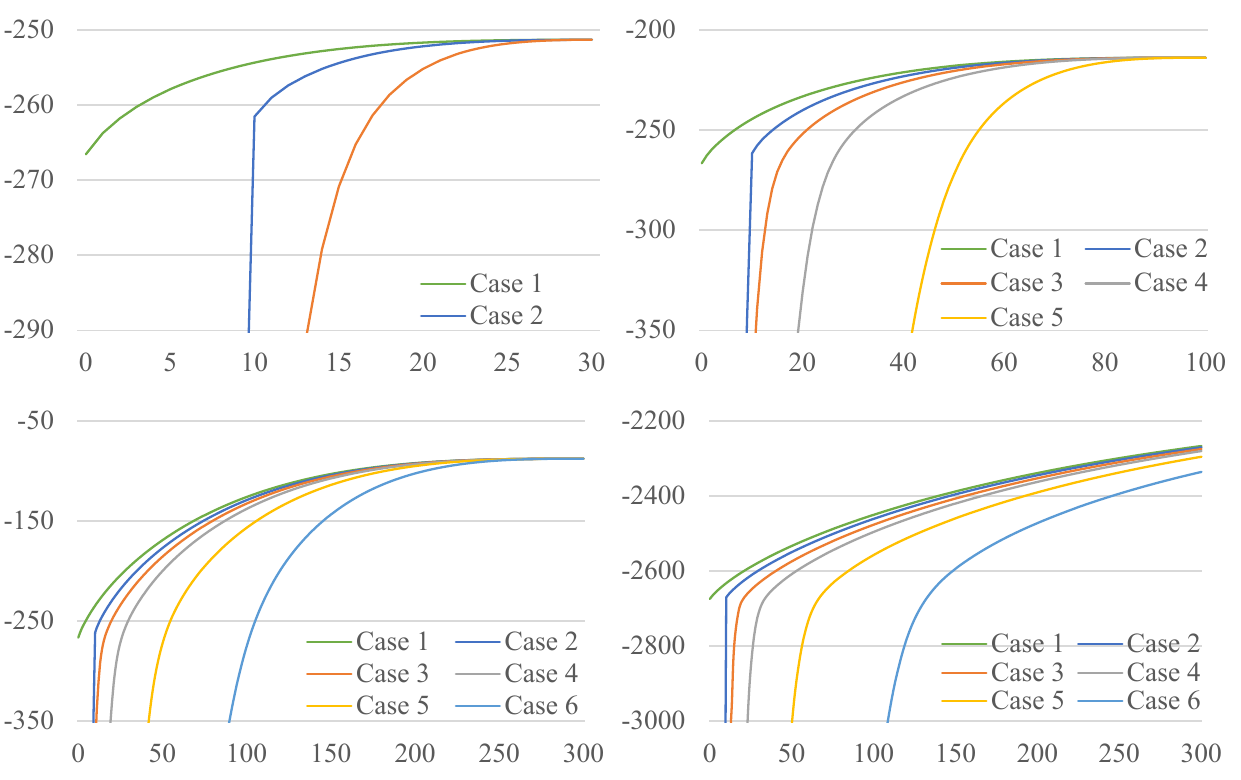}
\end{center}
\caption{Log-likehoods of various model orders for six cases. Left column: $m=1,\ldots ,30$, Right column: $m=1,\ldots ,k$. From top to bot, $k=30$, 100, 300 and 1000.}
\label{Fig_log-likelihood_Case1-6}
\end{figure}

Figure \ref{Fig_log-likelihood_Case1-6} shows the change in the maximum log-likelihood of the subset regression models for Cases 1--6, and $k=30$ (upper left), 100 (upper right), 1,000 (lower left) and 10,000 (lower right). The green curve shows the results for Case 1 where all regression coefficients are zero. In the subset regression model, the log-likelihood increases relatively rapidly at first as the order (shown in abscissa) increases, but hardly at all in the second half of the model orders. However, as already mentioned, since the true regression coefficients are zero, these are all spurious improvements, in reality, they are merely an accumulation of model errors.

The blue curve shows the case where only the first 10 coefficients are non-zero and all others are zero. In this case, the increase in log-likelihood for $0<m<10$ is very large. As shown in Figure \ref{Fig_summary_Case1-6}, the difference between the maximum log-likelihood and the expected log-likelihood is small in this order range, but when the number of data, $N$, is as large as 1000, the increase in log-likelihood is so rapid that adding a penalty equivalent to that in Case 1 has little effect on the order selection result. This suggests that by adding a penalty equivalent to that in Case 1, it is very likely that we may obtain good results in order selection.

In Case 3--6, all coefficients are non-zero, but their absolute values converge to zero, and their decay rates become progressively smaller. As the decay rate decreases, the log-likelihood continues to increase up to larger orders. This result suggests that when the decay rate of the true coefficients is small and a large number of coefficients are significant, the bias correction assuming that all coefficients are zero is not expected to be able to cope, and some measure is required.

\begin{figure}[tbp]
\begin{center}
\includegraphics[width=150mm,angle=0,clip=]{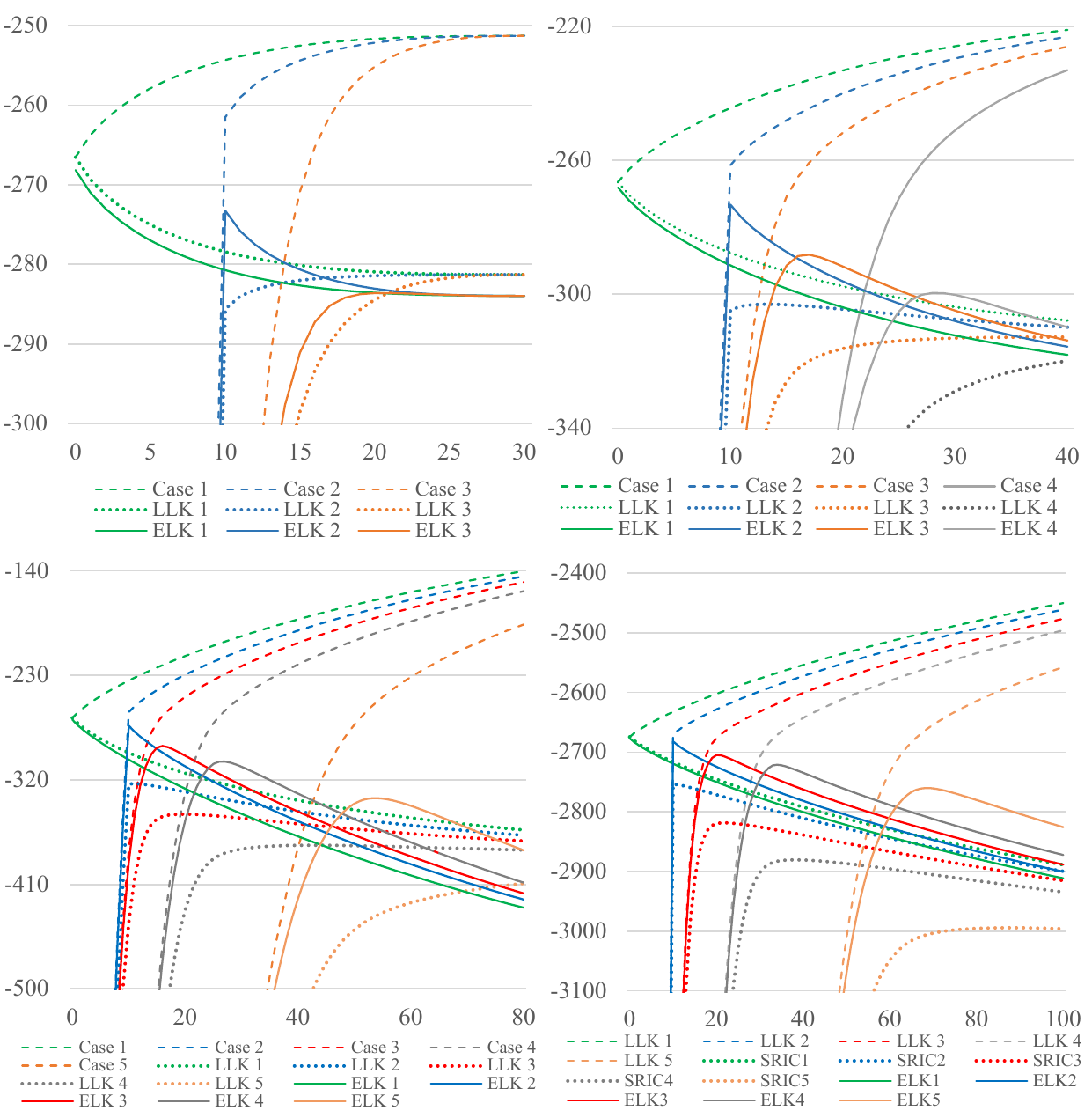}
\end{center}
\caption{Bias corrected of log-likehood by SRIC. Left column: $m=1,\ldots ,30$, Right column: $m=1,\ldots ,k$. From top to bot, $k=30$, 100, 300 and 1000.}
\label{Fig_Bias-correction_by_SRIC}
\end{figure}

Figure \ref{Fig_Bias-correction_by_SRIC} shows the change in log-likelihood when correcting for the bias theoretically obtained under the assumption of Case 1 for various cases, upper left, upper right, lower left, and lower right, in that order, for $k=30$, 100, 300, and 1000 candidate numbers of regression variables. The dashed line shows the log-likelihood before correction, the dotted line shows the log-likelihood after correction, and the solid line shows the expected log-likelihood. These results were obtained as an average of the results of 100,000 Monte Carlo experiments.

When all coefficients are zero, shown by the green line, and when only a few coefficients are significant, shown by the blue line, we can see that the appropriate order is all selected by correcting for log-likelihood. On the other hand, when all coefficients are non-zero (Cases 3-6), the results differ depending on the cases (i.e., different attenuation rates). For example, looking at the case $k=$1000, we can see that in Case 3 (red) and Case 4 (gray), the corrected log-likelihood reproduces well the trend of variation in the expected log-likelihood, indicating that the appropriate order can be selected. However, in Case 5 (orange), the correction is insufficient, resulting in the selection of a too high order. This result indicates that the correction assuming that all coefficients are zero is inappropriate when the ratio of significant variables is high relative to the number of explanatory variables.

\section{Adptive Bias Correction}

Based on the discussion in the previous section, this section considers how to modify the bias correction term to account for some patterns in the true regression model.
As shown in Figure \ref{Fig_log-likelihood_Case1-6}, in the subset regression model, when some of the true coefficients are significant, the log-likelihood increases rapidly at first, and the increase is much larger than the expected value of the extreme values of the chi-square variable.  Thus, if the increase in the log-likelihood is large enough, it can be considered not to be a spurious improvement due to error effects.

Here define the increase in the log-likelihood as
\begin{eqnarray}
 DF_m = \ell (\hat{\theta}_m) - \ell (\hat{\theta}_{m-1}),
\end{eqnarray}
and compare $DF_m$ and the expected value of the extreme values of the chi-squared variable in $k-m$ remaining explanatory variables, $X_{(1|k-m)}$.
If the expected value of the extreme value of the $\chi^2$ variable is greater than the increase in log-likelihood, $DF_m$, the increase in log-likelihood is likely to be due to error. On the other hand, if the opposite is the case, it is most likely due to an increase in the expressive power of the model due to an increase in the explanatory variables. However, it is not obvious that it is appropriate to switch when these two values become exactly the same, so the following quantities are considered:
\begin{eqnarray}
  \alpha_m = \frac{\exp\Bigl\{X_{(1|k-m)}\Bigr\}}{\exp\Bigl\{X_{(1|k-m)}\Bigr\} + \exp\Bigl\{DF_m\Bigr\}}.
\end{eqnarray}
This quantity represents the relative weight that the log-likelihood attributes to noise. Thus, the distribution of the point in time at which the noise-dominant coefficients begin is obtained by summing $\alpha_j$ starting with the first variable until the sum is 1, i.e.,
\begin{eqnarray}
  w_m = \min \left\{ \sum_{j=1}^m \alpha_j \, , \, 1 \right\}  .
\end{eqnarray}
 Then the penalty terms for each order of subset regression model is obtained as follows;
\begin{eqnarray}
  C_m = \sum_{j=1}^k w_j B_j^m ,
\end{eqnarray}
where $B_j^m$ is given by
\begin{eqnarray}
  B_j^m = \left\{  \begin{array}{ll} j  &  {}\quad\mbox{if }\, j < m \\
                                   m-1 + E[X_{(j-m|k-m)}] & {}\quad\mbox{if }\, j\geq m .
                 \end{array}\right.
\end{eqnarray}
Note that $E[X_{(j-m|k-m)}]$ is the expected value of the $(j-m)$-th ordered statistics for 
$k-m$ samples of $\chi^2$ random variable with one degree of freedom.

\begin{figure}[h]
\begin{center}
\includegraphics[width=160mm,angle=0,clip=]{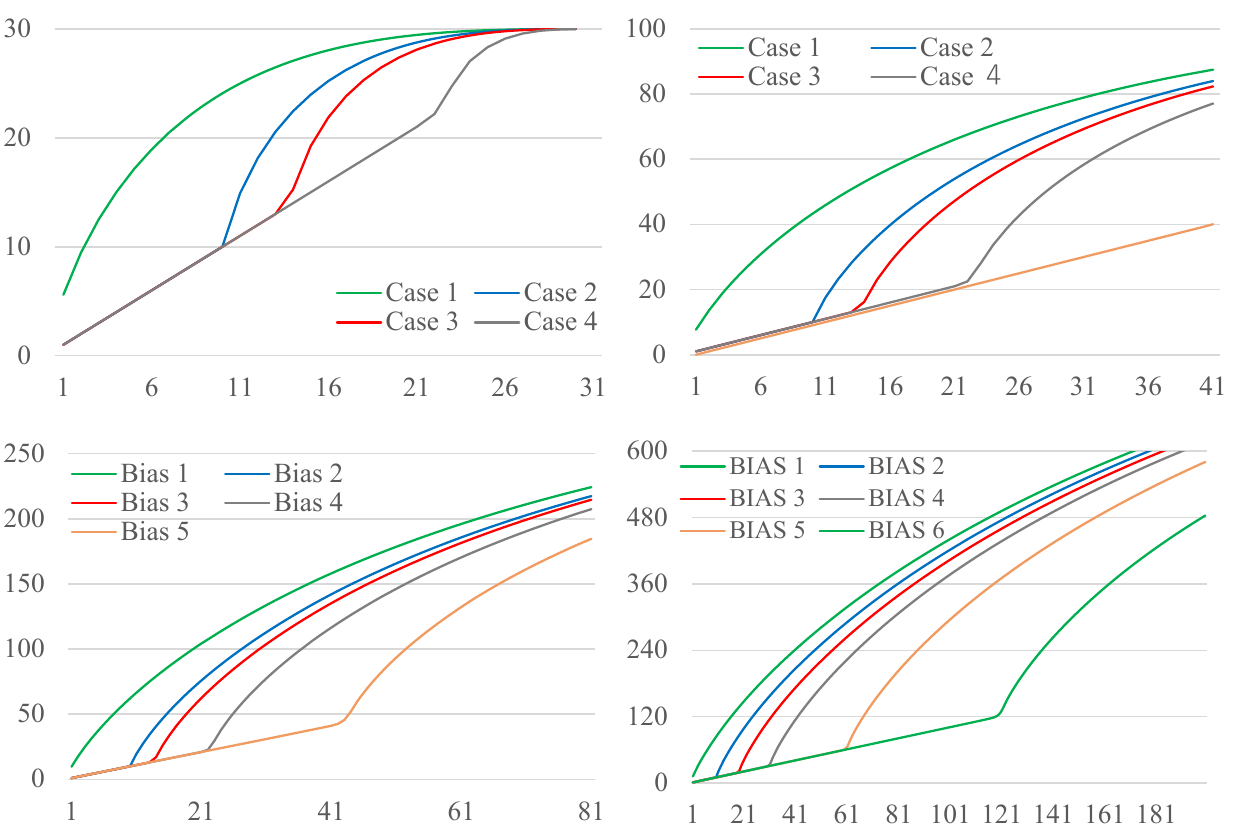}
\end{center}
\caption{Adaptive bias. Left column: $m=1,\ldots ,30$, Right column: $m=1,\ldots ,k$. From top to bot, $k=30$, 100, 300 and 1000.}
\label{Fig_Adaptive_bias}
\end{figure}

Figure \ref{Fig_Adaptive_bias} shows the bias estimates for $k=30$, 100, 300 and 1,000 obtained by this method.
The green curve is for Case 1, where all coefficients are assumed to be zero. As the order increases, the penalty increase becomes smaller and converges to zero. The blue line is for the Case 2 where the first 10 coefficients are non-zero; the slope is 1 up to the order 10 corresponding to the penalty of AIC, and thereafter the behavior is similar to Case 1 with $k-10$ variables.  For Cases 3--6, as the rate of decay of the true coefficients decreases, the interval of the straight line becomes longer. The smooth transition from a straight line to an upward curve is due to the distribution of $w_m$ over several points.

\begin{figure}[tbp]
\begin{center}
\includegraphics[width=160mm,angle=0,clip=]{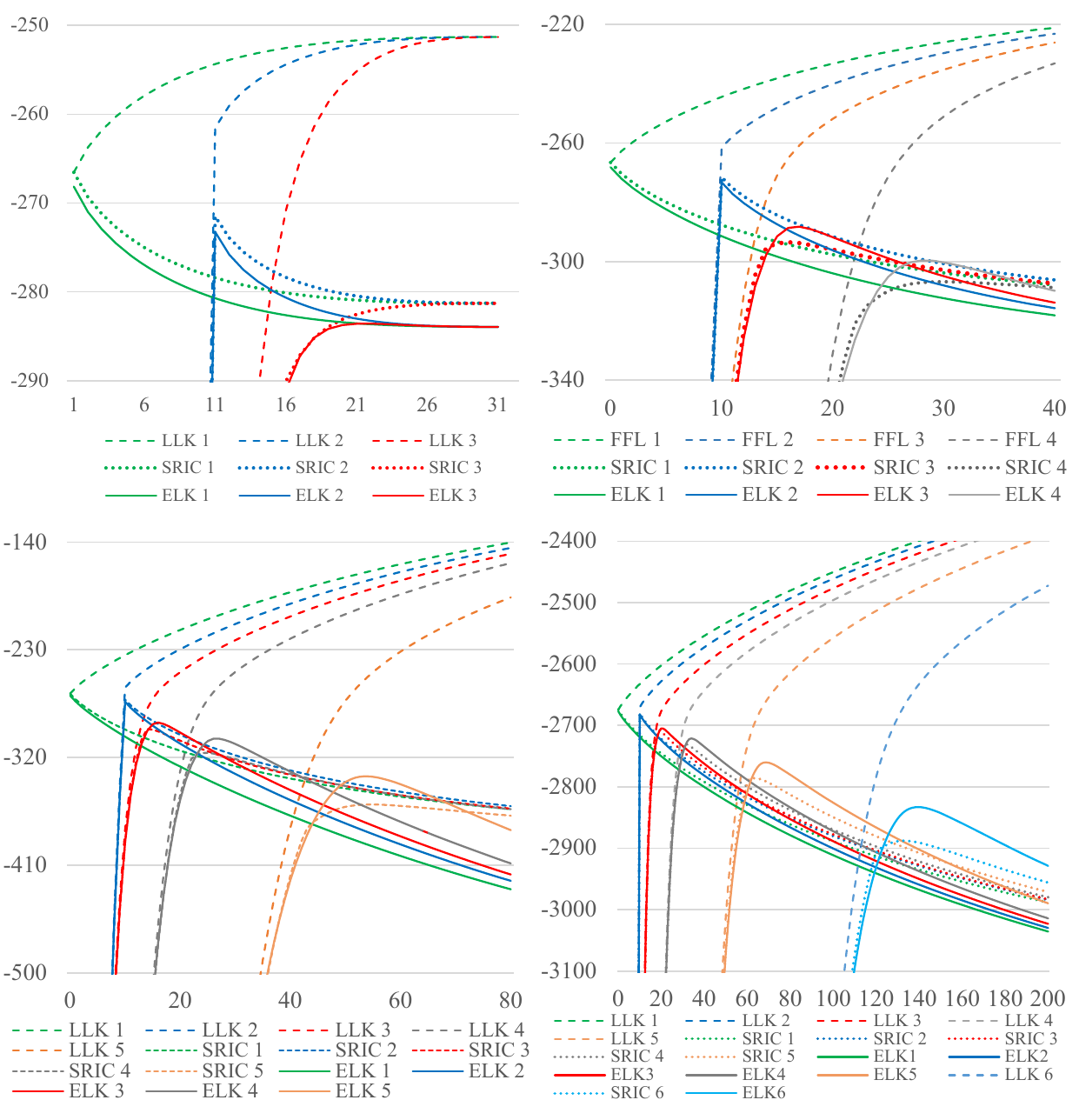}
\end{center}
\caption{Bias corrected log-likelihood by the adaptive bias. Left column: $m=1,\ldots ,30$, Right column: $m=1,\ldots ,k$. From top to bot, $k=30$, 100, 300 and 1000.}
\label{Fig_Bias_correction_by_adaptive_bias}
\end{figure}

Figure \ref{Fig_Bias_correction_by_adaptive_bias} shows the bias-corrected log-likelihood by this adaptive bias estimate. For $k=30$, Case 1 (green) and Case 2 (blue) show good corrections, selecting the order that maximizes the expected log-likelihood. The corrected log-likelihood in Case 3 (red) is a good approximation of the expected log-likelihood compared to the log-likelihood. It should be noted, however, that the order that maximizes the corrected log-likelihood is the maximum order (30), which is quite different from the optimal order 21 that maximizes the expected log likelihood.
At the same time, however, despite the too large order, the expected log-likelihood value is only 0.372, indicating that the model is not so bad in terms of expected log-likelihood.

For $k=100$, the corrected log-likelihood is a good approximation of the expected log-likelihood for all of Cases 1 through 4, and maximization of the corrected log-likelihood yields values close to the optimal order.
In Case 5 (orange) for $k=300$ and Case 6 (light blue) for $k=$1000, the maximum point of the curve is close to the order that maximizes the expected log-likelihood, but the kurtosis of the curve is smaller, suggesting that the approximation is not good when the significant coefficient is relatively large.

The above results show that the order selection of the subset regression model can be achieved relatively well by the bias-corrected log-likelihood. However, this bias-corrected order selection criterion was derived under the assumption that only a relatively small number of regression variables are valid. Since the expected value of the order statistic becomes almost zero as the order increases, it is likely that the correction term will not serve as a penalty and the maximum order will be selected if many of the candidate variables are needed. A safeguard in this case would be to also use the standard AIC, and adopt the lower of the two orders selected by the criteria proposed in this paper and the AIC.

\newpage
\noindent\textbf{\Large Appendix}\\[2mm]

\noindent\textbf{\Large A.1 \,Numerical Values of Figures \ref{Fig_ordered-chi-square_k=10,30,100,300,1000,3000,10000} and \ref{Fig_ordered-chi-square_k=100,1000}}
\begin{table}[h]
\caption{Expacted values of  $r$-th largest ($r=1,\ldots ,30$) $\chi^2$ distribution for various values of number of samples, $k$=30, 100, 300, 1,000, 3,000 and 10,000.}\label{Tab_Expacted values of ordered chi-square}
\begin{center}
\tabcolsep=1.5mm
\begin{small}
\begin{tabular}{r|c|rrrrrrr}
    &               & \multicolumn{7}{c}{ $k$ } \\
$r$ & $E[\chi^2_1]$ & $10$ & $30$ & $100$ & $300$ & $1,000$ & $3,000$ & $10,000$    \\
\hline
   1  & 1 & 3.799 & 5.599 & 7.703 & 9.686 &11.889 &13.892 &15.988\\
   2  & 1 & 2.171 & 3.868 & 5.911 & 7.862 &10.055 &12.089 &14.344\\
   3  & 1 & 1.426 & 3.038 & 5.034 & 6.960 & 9.135 &11.158 &13.403\\
   4  & 1 & 0.971 & 2.502 & 4.458 & 6.364 & 8.526 &10.540 &12.778\\
   5  & 1 & 0.660 & 2.113 & 4.033 & 5.921 & 8.071 &10.078 &12.310\\
   6  & 1 & 0.438 & 1.810 & 3.697 & 5.570 & 7.709 & 9.709 &11.937\\
   7  & 1 & 0.275 & 1.565 & 3.419 & 5.278 & 7.408 & 9.403 &11.627\\
   8  & 1 & 0.158 & 1.361 & 3.184 & 5.030 & 7.151 & 9.141 &11.361\\
   9  & 1 & 0.076 & 1.187 & 2.981 & 4.814 & 6.927 & 8.913 &11.129\\
  10  & 1 & 0.025 & 1.038 & 2.801 & 4.622 & 6.729 & 8.710 &10.923\\
  11  & 1 & 0.000 & 0.908 & 2.642 & 4.451 & 6.550 & 8.528 &10.738\\
  12  & 1 & 0.000 & 0.793 & 2.498 & 4.296 & 6.389 & 8.362 &10.569\\
  13  & 1 & 0.000 & 0.692 & 2.367 & 4.154 & 6.241 & 8.211 &10.415\\
  14  & 1 & 0.000 & 0.602 & 2.247 & 4.024 & 6.105 & 8.071 &10.273\\
  15  & 1 & 0.000 & 0.521 & 2.137 & 3.904 & 5.979 & 7.942 &10.142\\
  16  & 1 & 0.000 & 0.449 & 2.035 & 3.792 & 5.861 & 7.821 &10.019\\
  17  & 1 & 0.000 & 0.385 & 1.940 & 3.687 & 5.751 & 7.708 & 9.904\\
  18  & 1 & 0.000 & 0.327 & 1.852 & 3.589 & 5.648 & 7.602 & 9.795\\
  19  & 1 & 0.000 & 0.276 & 1.769 & 3.497 & 5.550 & 7.502 & 9.693\\
  20  & 1 & 0.000 & 0.230 & 1.692 & 3.410 & 5.458 & 7.407 & 9.597\\
  21  & 1 & 0.000 & 0.189 & 1.619 & 3.328 & 5.371 & 7.317 & 9.505\\
  22  & 1 & 0.000 & 0.152 & 1.549 & 3.250 & 5.288 & 7.232 & 9.417\\
  23  & 1 & 0.000 & 0.120 & 1.484 & 3.175 & 5.209 & 7.150 & 9.334\\
  24  & 1 & 0.000 & 0.093 & 1.422 & 3.104 & 5.133 & 7.072 & 9.254\\
  25  & 1 & 0.000 & 0.069 & 1.363 & 3.036 & 5.061 & 6.998 & 9.178\\
  26  & 1 & 0.000 & 0.049 & 1.307 & 2.972 & 4.992 & 6.926 & 9.105\\
  27  & 1 & 0.000 & 0.032 & 1.254 & 2.909 & 4.925 & 6.857 & 9.035\\
  28  & 1 & 0.000 & 0.019 & 1.203 & 2.850 & 4.861 & 6.791 & 8.967\\
  29  & 1 & 0.000 & 0.009 & 1.154 & 2.792 & 4.800 & 6.727 & 8.902\\
  30  & 1 & 0.000 & 0.003 & 1.108 & 2.737 & 4.740 & 6.666 & 8.839\\
\hline
\end{tabular}\end{small}
\end{center}
\end{table}

\noindent\textbf{\Large A.2 \,Note on the Expectation of Order Statistic of $\chi^2_m$ Distribution}\\[2mm]

In this paper, the explanatory variables are assumed to be orthogonal, so this is not directly necessary, but a note on the expected value of the order statistic of the chi-square variable with m degrees of freedom is provided for the general case.

Assume that $X_n, n=1,\ldots ,N$ is distributed according to the density function $f(x)$.
Here, we define the order statistics in order of magnitude as follows;  
\begin{eqnarray}
  X_{(1)} \leq X_{(2)} \leq \ldots \leq X_{(N)}.
\end{eqnarray}
Then the distribution of the order statistics is given by
\begin{eqnarray}
  p_{x_{(r)}}(x) = \frac{N!}{(r-1)!(N-r)!}F(x)^{r-1}\{1-F(x)\}^{N-r}f(x).
\end{eqnarray}
Distribution of two order statistics, $X_{(r)},X_{(s)}, r<s$: 
\begin{eqnarray}
  \lefteqn{ p_{x_{(r)},x_{(s)}}(x,y)} \nonumber \\
  &=& \frac{N!}{(r-1)!(s-1)!(N-s)!}F(x)^{r-1}\{F(y)-F(x)\}^{s-r-1}\{1-F(y)\}^{N-s}f(x)f(y).
\end{eqnarray}
Distribution of $k$ order statistics, $X_{(r_1)},\ldots,X_{(r_k)}, r_1<\ldots <r_k$:
\begin{eqnarray}
  \lefteqn{ p_{x_{(r_1)},\ldots ,x_{(r_k)}}(x_1,\ldots ,x_k) 
      = \frac{N!}{(r_1-1)!(r_2-r_1-1)!\cdots (N-r_k)!} } \nonumber \\
  && \times F(x_1)^{r_1-1}\{F(x_2)-F(x_1)\}^{r_2-r_1-1}\cdots \{1-F(x_k)\}^{N-r_k} f(x_1) \cdots f(x_k).
\end{eqnarray}
Distribution of all ordered statistics, $X_{(1)},\ldots,X_{(N)}$
\begin{eqnarray}
  p_{x_{1},\ldots ,x_{N}}(x_1,\ldots ,x_N) = N! f(x_1)\times \cdots\times f(x_N).
\end{eqnarray}

\vspace{3mm}
\textbf{$\chi^2$ distribution with the degree of freedom $k$}\\
Density function;
\begin{eqnarray}
f(x;k) = \frac{1}{2^{k/2}\Gamma (\frac{k}{2})}x^{\frac{k}{2}-1}\exp^{-\frac{x}{2}}.
\end{eqnarray}
Distribution function;
\begin{eqnarray}
F(x;k) = \frac{\gamma(\frac{k}{2},\frac{x}{2})}{\Gamma (\frac{k}{2})}.
\end{eqnarray}
where $\gamma(\frac{k}{2},\frac{x}{2})$ : Incomplete Gamma function;
\begin{eqnarray}
\gamma\left(k,x\right) &=& (k-1) \gamma\left(k-1,x\right) - x^{k-1} \exp^{-x} \\
\gamma(1,x)           &=& 1 - \exp^{-x} \nonumber \\
\gamma(1/2,x) &=& \sqrt{\pi} \mbox{erf}(\sqrt{x}) \nonumber \\
\gamma\left(\frac{k}{2},\frac{x}{2}\right) &=& \left(\frac{k}{2}-1\right) \gamma\left(\frac{k}{2}-1,\frac{x}{2}\right) - \left(\frac{x}{2}\right)^{\frac{k}{2}-1} \exp^{-\frac{x}{2}} .\nonumber
\end{eqnarray}
Using the recurrence relation and the initial values, the incomplete Gamma function is given by

For $k$ even number:
\begin{small}
\begin{eqnarray}
\gamma\left(\frac{k}{2},\frac{x}{2}\right) 
  &=& \left(\frac{k}{2}-1\right)! - \exp^{-\frac{x}{2}} \left\{\left(\frac{k}{2}-1\right)! + \sum_{j=1}^{\frac{k}{2}-1} \left(\frac{k}{2}-j\right)!\left(\frac{x}{2}\right)^j \right\}  \\
  &=& \left(\frac{k}{2}-1\right)! - \left\{\left(\frac{k}{2}-1\right)! + \left(\frac{k}{2}-1\right)!\left(\frac{x}{2}\right) + \cdots + \left(\frac{x}{2}\right)^{k/2-1} \right\} \exp^{-\frac{x}{2}} .\nonumber
\end{eqnarray}
\end{small}
%

For $k$ odd number:
\begin{eqnarray}
\gamma\left(\frac{k}{2},\frac{x}{2}\right) 
  &=& \frac{(k-2)!!}{2^{(k-1)/2}}\sqrt{\pi} \mbox{erf}(\sqrt{x}) 
- \exp^{-\frac{x}{2}} \sum_{j=1}^{\frac{k-1}{2}}\frac{(k-2)!!}{2^{(\frac{k-1}{2}-j)}(2j-1)!!} \left(\frac{x}{2}\right)^{j-\frac{1}{2}} \\
  &=& \frac{((k-1)/2)!!}{2^{(k-1)/2}} - \left\{\frac{(k-2)!!}{2^{(k-1)/2}}\left(\frac{x}{2}\right) + \frac{(k-4)!!}{2^{(k-3)/2}}\left(\frac{x}{2}\right)^2 + \cdots + \left(\frac{x}{2}\right)^{k/2-1} \right\} \exp^{-\frac{x}{2}}. \nonumber
\end{eqnarray}
%

\end{document}